\documentclass{PoS}
\usepackage{bm}
\title{Dark matter search in the inner Galactic halo with H.E.S.S. I and H.E.S.S. II}

\ShortTitle{Dark matter in the inner GC halo with H.E.S.S.}

\author{\speaker{V. Lefranc$^{1}$} and E. Moulin$^{1}$, for the H.E.S.S. collaboration\\
        E-mail: \email{valentin.lefranc@cea.fr}, \email{emmanuel.moulin@cea.fr}\\
     $^1$ DSM/Irfu, Service de Physique des Particules, CEA Saclay, F-91191 Gif-Sur-Yvette Cedex, France\\
     }     
\abstract{The presence of dark matter in the Universe is nowadays widely supported by a large body of astronomical and cosmological observations. One of the best targets to look for dark matter particle self-annihilation into very high energy gamma-rays is the Galactic Center (GC) region. A search for annihilating dark matter in the central 300 parsecs around the GC is performed with the H.E.S.S. array of ground-based Cherenkov telescopes. Using the full 
H.E.S.S.- I dataset (2004-2014) for this region, new constraints are derived on the velocity-weighted annihilation cross section with a 2D likelihood method, taking advantage of differences in both spectral and spatial morphologies of signal and background. Higher statistics from the 10-years GC dataset of H.E.S.S. I together with a novel analysis technique, allow to improve the present constraints by a factor of 2 for a DM particle mass of 1 TeV. 
The expected improvement in sensitivity of the new analysis technique, applied to performances of the H.E.S.S.-II array, 
is also presented.}

\FullConference{The 34th International Cosmic Ray Conference,\\
		30 July- 6 August, 2015\\
		The Hague, The Netherlands}

\begin{document}
\section{Introduction}
Dark Matter (DM) constitutes  about 85\% of the mass content of the Universe. Its existence is supported by a wealth of astrophysical  and cosmological measurements. There are many well-motivated elementary particle candidates, arising in extensions of the Standard Model of particle physics. Some of the most compelling classes of models assume DM to consist  of  Weakly Interactive Massive Particles (WIMPs): stable particles with masses and coupling strengths at the electroweak scale produced in a standard thermal history of the Universe have the relic density that corresponds to that of observed DM. DM particles would then self-annihilate into Standard Model particles, including gamma-rays, neutrinos and charged particles in the final state. The indirect searches look for the detection of these secondary products, which provides unique test of the DM nature in the cosmos and provide the necessary link between DM in the Universe and  discoveries at terrestrial accelerators.

The expected annihilation rate is proportional to the square of the DM density, integrated along the line of sight (l.o.s). Regions with enhanced DM density are therefore primary targets for indirect searches.
The Galactic Center (GC) is one of the most promising observational targets due to its proximity and high DM content. 
Currently, the strongest constraints on the velocity-weighted annihilation cross section for DM particles in the TeV mass range are obtained from the ground-based array of Cherenkov telescopes,  H.E.S.S., reaching an upper limit at $\rm{3 \times 10^{-25}\ cm^{3}s^{-1}}$, using 112 hours of observation towards the GC region~\cite{Abramowski:2011hc}. 
Here, we perform a new analysis of the inner GC region with the 10-year GC dataset of H.E.S.S. I (only the four small telescopes)
using both the spectral and spatial characteristics of the DM annihilation signal. 

Since September 2012,  the phase II of H.E.S.S. started with the addition of a large telescope at the centre of the existing array, extending the accessible DM mass range down to several tens of GeV. The expected sensitivity of the H.E.S.S.-II array for the inner GC halo will also be presented.

\section{Dark Matter search with H.E.S.S. in the Galactic halo}
\subsection{Galactic Center observations}
The High Energy Stereoscopic System (H.E.S.S.), operating since 2003, is an array of Imaging Atmospheric  Cherenkov  telescopes located in the Khomas Highland of Namibia operating since 2003~\cite{hess}. The H.E.S.S. instrument detects very-high-energy $\gamma$-rays (E $\gtrsim$ 100 GeV) by collecting the Cherenkov light produced by electromagnetic showers induced by a primary 
gamma-ray interacting in the Earth's upper atmosphere. %Cosmic rays (CRs) such as protons and electrons are primary sources of background.  the showers they initiate can mimick gamma-ray showers and constitue the irreducible background. 
The H.E.S.S. array is made of 4 telescopes with a dish of 12 m  diameter with a total field of view (FoV) of $5^{\circ}$, and a larger one of 28 m  diameter located at the center of the array aiming at lowering the energy threshold down to a few tens of GeV, with a FoV of $3.5^{\circ}$.

\begin{figure}[t]
\centering
\includegraphics[width=0.8\textwidth]{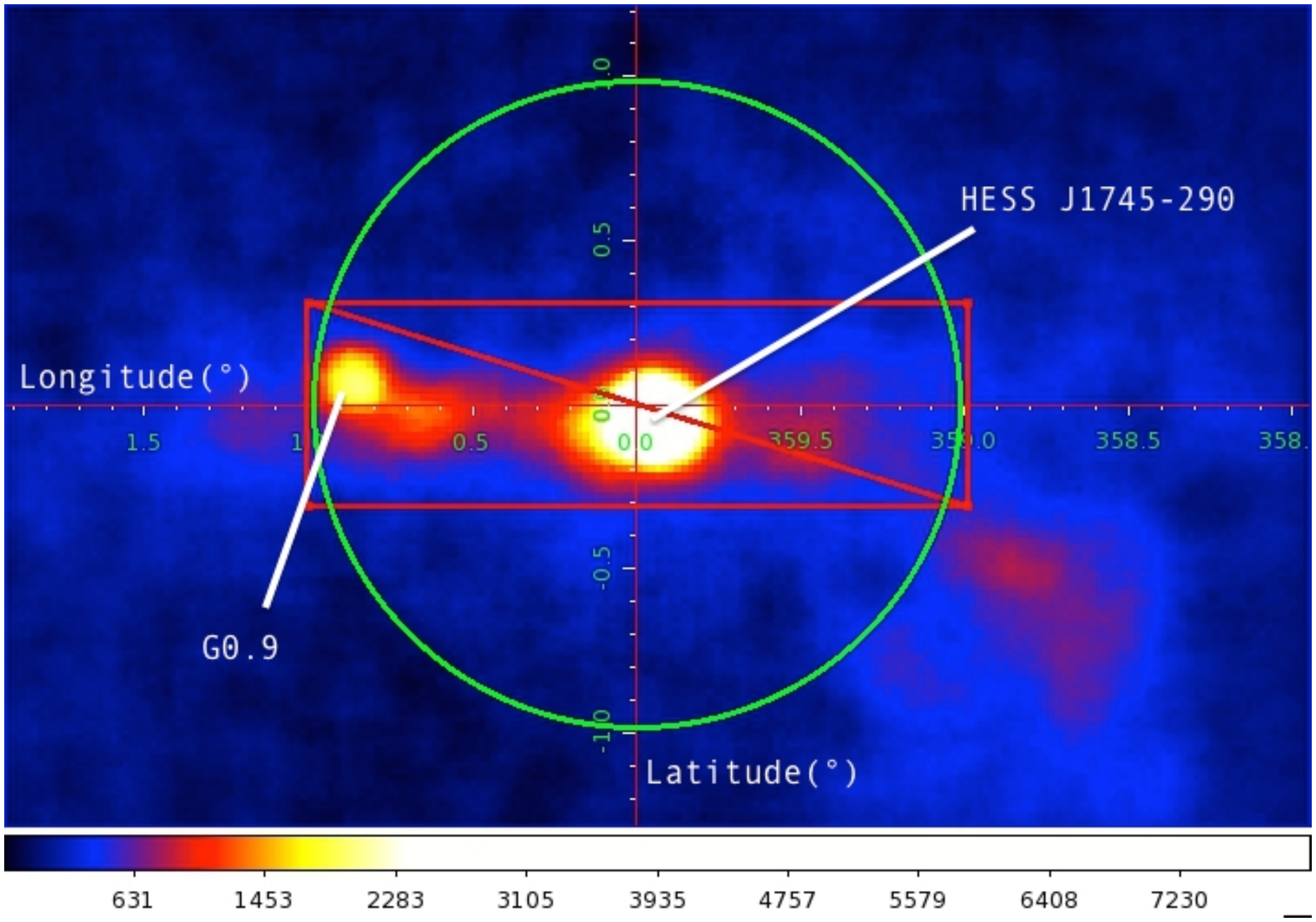}\
\caption{The Galactic Center region viewed by the  phase-I H.E.S.S. instrument: the sky map shows the gamma-ray excess map in Galactic coordinates from 254 live hours. The RoI is shown as a green circle from which Galactic latitudes 
|b|$<$0.3$^{\circ}$ (red box) are excluded to avoid background contamination from detected sources~\cite{Aharonian:2004wa,Aharonian:2005br,Aharonian:2006pe,Aharonian:2009zk} and diffuse emission~\cite{Aharonian:2006au}. 
} 
\label{fig:GC}
\end{figure}

The data analysis is carried out using 254 live hours coming from the full 2004-2014 dataset of GC observations with H.E.S.S. phase I (see Fig.~\ref{fig:GC}). The run selection follows the H.E.S.S. standard quality 
criteria~\cite{Aharonian:2006pe}. 
% and an additional one on the zenith angle of observations at 45$^{\circ}$ is used to avoid possible systematics. 
The mean zenith angle of observations for the dataset is $18^{\circ}$. Data analysis is performed on a region of interest (RoI),
given by a circular region of $1^{\circ}$ around the GC, hereafter referred to as the ON region. Galactic latitudes of $\rm{|b|<0.3^{\circ}}$ are excluded in order to avoid contamination from standard astrophysical
emissions~\cite{Aharonian:2004wa,Aharonian:2005br,Aharonian:2006pe,Aharonian:2006au,Aharonian:2009zk}. The residual background is from an annulus of inner and outer radii of  $1^{\circ}$ and 1.5$^{\circ}$, respectively, 
 hereafter referred to as the OFF region. The normalization of the background accounts for acceptance gradients between the ON and OFF regions.  %The amount of dark matter in this OFF region is non zero but significantly smaller compared to the ON region due to the considered Einasto profile (see section 4).

\subsection{Dark matter annihilation flux}
The differential  $\gamma$-ray flux due to annihilation of self-conjugated DM particles of mass $\rm{m_{DM}}$ in a solid angle $\rm{d \Omega}$, is given by:
\begin{equation}
\label{promptflux}
\frac{\rm d \Phi_\gamma^{\rm P}}{\rm d \Omega d E_{\gamma}} = \frac{1}{8\pi}{m_{\rm DM}^2} J(\theta) \sum_f \langle { \sigma v} \rangle_f \frac{{\rm d} N^f_\gamma}{{\rm d} E_{\gamma}}(E_{\gamma}) \ , \qquad J(\theta)=\int_{\rm l.o.s.}{\rm d} s \ \rho^2(r(s,\theta)) \,
\end{equation}
where  $\langle { \sigma v} \rangle_f$ is  the thermally-averaged velocity-weighted annihilation cross section and 
${\rm d}  N^f_\gamma / {\rm d} E_{\gamma}$  is the energy spectrum of photons per annihilation in the channel with final state $f$, respectively.  

The coordinate $r$ is measured from the GC, and can expressed as  $r(s, \theta)=(r_\odot^2+s^2-2 r_\odot s \cos\theta)^{1/2}$, where $s$ is the distance along the line of sight (l.o.s.) and  $\theta$ is the angle between the l.o.s. and the direction towards the GC, and $r_\odot=8.5$ kpc is the distance of the Sun ot the GC. The function $J(\theta)$, commonly referred to as the {\it J-factor}, is the l.o.s. integral of  the square of the DM density $\rho$. In this analysis,  this density is assumed to follow an Einasto profile, parametrized by:
\begin{equation}
\rho(r) = \rho_{\rm s}  \exp \left[-\frac{2}{\alpha_{\rm s}}\left(\Big(\frac{r}{r_{\rm s}}\Big)^{\alpha_{\rm s} }-1\right)\right] \ . 
\end{equation}
The parameters ($\rho_{\rm s}, \alpha_{\rm s}, r_{\rm s}$) are extracted from Ref.~\cite{Abramowski:2011hc}.  The $\gamma$-ray spectrum from DM annihilation in a channel $f$ is computed by using the tools available from Ref.~\cite{Cirelli:2010xx}.

\subsection{Analysis methodology}
The statistical method used in this analysis is based on a likelihood ratio test which 
takes full advantage of the differences between the spatial and spectral features of the DM signal and the background. 
The RoI is divided into 7 spatial bins defined as annuli of $0.1^{\circ}$ width, from $0.3^{\circ}$ to $1^{\circ}$ in radial distance from the GC. 
The background is estimated from an annulus outside the RoI, with inner and outer radii of $1^{\circ}$ and $1.5^{\circ}$, respectively.
For a given DM mass $m_{DM}$, the total likelihood is obtained by the product over the spatial bins $i$ and the energy bins $j$ of the individual Poisson likelihoods. It explicitly writes:
\begin{equation}
 \mathcal{L}  (m_{\rm DM}, \langle { \sigma v} \rangle ) = \prod_{\rm i,j} \left( \mathcal{L}_{\rm ij}^{\rm ON}   (m_{\rm DM},\langle { \sigma v} \rangle) \right) \times \prod_{\rm j} \left(  \mathcal{L}_{\rm j}^{\rm OFF}    (m_{\rm DM},\langle { \sigma v}\rangle) \right)
 \ ,
\end{equation}
where the spectral part of the likelihood covers the energy range from threshold up to $m_{\rm DM}$, while  the spatial part covers the seven sub-RoIs. Following Ref.~\cite{2011EPJC711554C}, the individual likelihoods are given by
\begin{equation}\label{Likelihood}
\mathcal{L}_{\rm ij}^{\rm ON}(\bm{N^{\rm S_{ON}},N^{\rm B}|N_{\rm ON},N_{\rm OFF}}) = \frac{\left(N_{\rm ij}^{\rm S_{\rm ON}}+\alpha_i N_{\rm j}^{\rm B}\right)^{N_{{\rm ON}, {\rm ij}}}}{N_{{\rm ON}, {\rm ij}}!}e^{-(N_{\rm ij}^{\rm S}+ \alpha_i N_{\rm j}^{\rm B})}   \ 
\end{equation}
\begin{equation}\label{LikelihoodOFF}
\mathcal{L}_{\rm  j}^{\rm OFF}(\bm{ N^{\rm S_{OFF}},N^{\rm B}|N_{\rm ON},N_{\rm OFF}}) = \frac{\left(N_{\rm j}^{\rm B}+N_{\rm j}^{\rm S_{\rm OFF}}  \right)^{N_{{\rm OFF}, {\rm j}}}}{N_{{\rm OFF},{\rm j}}!} e^{-(N_{\rm j}^{\rm B}+N_{\rm j}^{\rm S_{OFF}})} \ ,
\end{equation}
where $N_{ij}^{\rm S_{ON}}$ and $N_{j}^{\rm S_{OFF}}$ are the predicted annihilation signals in the spatial bin $i$ and spectral bin $j$  in the ON and OFF regions, respectively, $N_{ij}^{\rm S_{ON}}+\alpha_i N_{j}^{\rm B}$ is the expected total number of events in  the ON region,
$N_{{\rm ON},ij}$ stands for the total number of observed events in the ON region, and  $N_{{\rm OFF}, j}$ 
is the total number of events measured in the OFF region. $\bm{ N^{\rm S_{ON}}, N^{\rm S_{OFF}},N^{\rm B},N_{\rm ON}}$ and $\bm{ N_{\rm OFF}}$ are the vectors corresponding to the number of events prevouisly defined. The parameter $\alpha_i=\Delta\Omega_i/\Delta\Omega_{\rm OFF}$ refers to the ratio between the angular size of the ON region $i$ and the OFF region. 

Constraints on $\langle{ \sigma v}\rangle$ are obtained from the likelihood ratio test statistic given by:
 \begin{equation} 
 {\rm TS}=-2 \ln(\mathcal{L}(m_{DM},\langle { \sigma v} \rangle)/\mathcal{L}_{\rm max}(m_{DM},\langle { \sigma v} \rangle)) \ ,
\end{equation}
  which follows an approximate $\chi^2$ distribution with one degree of freedom. Values of $\langle { \sigma v}\rangle$ for which TS is higher than 2.71 are excluded at 95\% Confidence Level (C.L.).

\section{Results}
\begin{figure}[t]
\centering
\includegraphics[width=1\textwidth]{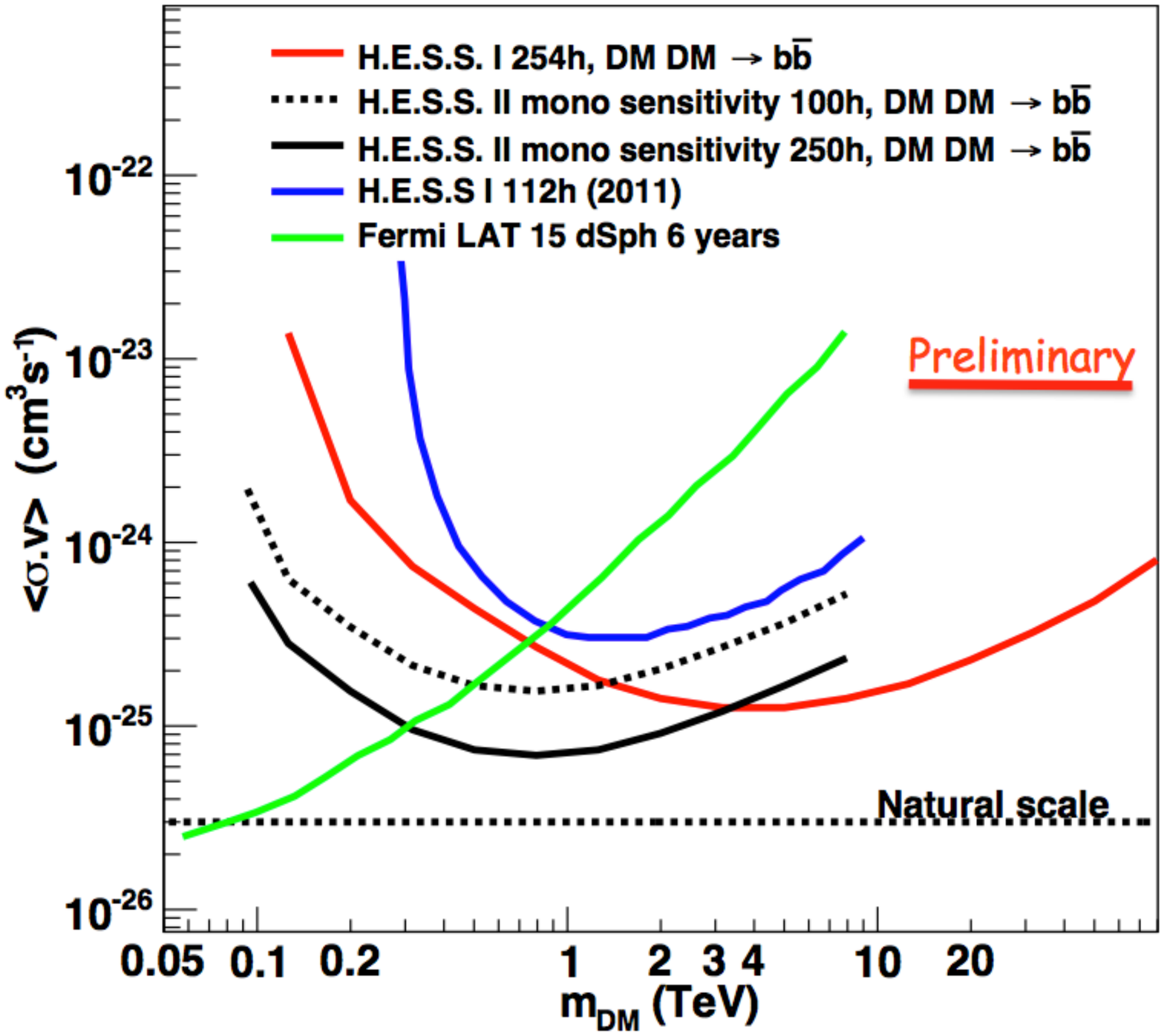}
\caption{Constraints on the thermally-averaged velocity-weighted cross section $\langle \sigma v \rangle$ for DM particle self-annihilation into $b\bar{b}$-pairs from the full H.E.S.S.-I dataset for the GC (red solid line). The previously reported H.E.S.S.-I limits are shown as blue solid line. The most recent Fermi-LAT constraints from observations of close-by dwarf spheroidal 
galaxies~\cite{Ackermann:2015zua} are given by the solid green line. The  H.E.S.S.-II  sensitivity  in mono mode is shown for 100  hours (dashed black line) and 250 hours (solid black line) of observations, respectively. The  expectation from the
DM relic density (natural scale), $\langle { \sigma v} \rangle = 3 \times 10^{-26}$ cm$^3$s$^{-1}$, is also plotted (dashed black line).}
\label{fig:results}
\end{figure}

No significant $\gamma$-ray excess is found in any of the RoIs considered here.
%No significant possible contribution comming from dark matter has been detected in the considered RoI. 
We thus derive upper limits on the velocity-weighted annihilation cross section.
Figure~\ref{fig:results} shows the 95\% C.L. upper limits on $\langle { \sigma v} \rangle$ versus the DM mass, assuming that the DM particles self-annihilate exclusively into $\rm{b\bar{b}}$-pairs. The strongest constraint is obtained for a  3 TeV DM mass and yields $\langle { \sigma v} \rangle$ $<$ $1.2 \times 10^{-25}$ cm$^3$s$^{-1}$.  We obtain stronger limits compared to the ones previously reported by H.E.S.S.  in Ref.~\cite{Abramowski:2011hc}, both at low and high DM masses. This improvement is mainly due to the novel analysis technique employed here which takes into account both the spectral and spatial behavior of the DM signal and the background, higher statistics from the full H.E.S.S.-I dataset for the GC, and up-to-date DM annihilation spectra~\cite{Cirelli:2010xx}. For comparison, we show the constraints obtained towards 15 dwarf spheroidal galaxies for 6 years of observations with Fermi-LAT~\cite{Ackermann:2015zua}. The H.E.S.S. limits well complement Fermi-LAT limits above $\sim$800 GeV, and provide the strongest limits for TeV DM particles to date.  

%  which gives stronguest constraints at high energy. (ii) we perform a 2D-likelihood method which improves the previous limits a factor $\sqrt{2}$ thanks to the different behaviour of the signal compare to the background in both spatial and spectral bins. Finally we gain another factor  $\sim \sqrt{2}$ by increase the observation time of a factor $\sim 2$ compare to previous study.
A sensitivity limit for H.E.S.S. II in mono mode\footnote{The mono mode corresponds to data taking with the 28-meter telescope only.}, using the same analysis methodology as for the H.E.S.S.-I analysis, is also shown in Fig.~\ref{fig:results} for 100 and 250 hours of observation, reaching maximal sensitivities in $\langle { \sigma v} \rangle$ of 1.5 $\rm \times 10^{-25}$ cm$^3$s$^{-1}$ and $7 \times 10^{-26}$ cm$^3$s$^{-1}$ at 800 GeV  respectively. This shows the potential of H.E.S.S. II to significantly improve the limits of H.E.S.S.-I below 1 TeV.

\section{Summary}
We performed a new search for a signal of self-annihilating DM particles in the inner Galactic halo.  We made use the full dataset of H.E.S.S.-I available from 10 years of observations of the GC, which correspond to 254 hours of live time. In this dataset, no significant gamma-ray excess was found in any of the considered RoIs.  We derived 95\% C.L. upper limits on 
$\langle { \sigma v} \rangle$ for DM particles annihilating in the $\rm{b\bar{b}}$ channelwith a 100$\%$ branching ratio. Together with higher statistics and a novel statistical analysis method we improved the limits down to $\rm{1.2 \times 10^{-25}}$ cm$^3$s$^{-1}$ for a DM mass of 4 TeV . This is a factor of 2.3 better than the strongest constraint obtained in the previous analyses, allowing to get closer to the scale of 3 $\times\ 10^{-26}$ cm$^3$s$^{-1}$, expected for the thermal DM relic density. The new result of H.E.S.S. on DM searches sets the strongest constraints on annihilating DM so far above$~\sim 800$ GeV. The H.E.S.S.-II sensitivity in mono mode shows that improvements are expected in the coming years due to a lower energy threshold, allowing H.E.S.S.-II to be competitive with Fermi-LAT down to a few hundred GeV.

%\begin{thebibliography}{99}
%\bibitem{...} 
%....
\bibliographystyle{ieeetr}
\bibliography{skeleton}
%\end{thebibliography}

\end{document}